\documentclass[aps,prl,twocolumn,footinbib,superscriptaddress,longbibliography,amsmath,amssymb]{revtex4-2} 

\usepackage{graphicx}
\usepackage{subfigure}
\usepackage{epsfig} 
\usepackage{dcolumn}
\usepackage{bm}
\usepackage{times} 
\usepackage{amsmath}
\usepackage{float}
\usepackage{color}
\usepackage{multirow}
\usepackage[normalem]{ulem}

\begin{document}
\title{Controlling charge and spin transport in an Ising-superconductor Josephson junction}

\author{Gaomin Tang}
\affiliation{Department of Physics, University of Basel, Klingelbergstrasse 82, CH-4056
Basel, Switzerland}
\author{Raffael L. Klees}
\affiliation{Fachbereich Physik, Universit\"{a}t Konstanz, D-78457 Konstanz, Germany}
\author{Christoph Bruder}
\affiliation{Department of Physics, University of Basel, Klingelbergstrasse 82, CH-4056
Basel, Switzerland}
\author{Wolfgang Belzig}
\affiliation{Fachbereich Physik, Universit\"{a}t Konstanz, D-78457 Konstanz, Germany}

\begin{abstract}
  An in-plane magnetic field applied to an Ising superconductor converts spin-singlet
  Cooper pairs to spin-triplet ones. In this work, we study a Josephson junction formed by
  two Ising superconductors that are proximitized by ferromagnetic layers. 
  This leads to highly tunable spin-triplet pairing correlations which allow to modulate
  the charge and spin supercurrents through the in-plane magnetic exchange fields. For a
  junction with a nonmagnetic barrier, the charge current is switchable by changing the
  relative alignment of the in-plane exchange fields, and a $\pi$-state can be realized.
  Furthermore, the charge and spin current-phase relations display a $\phi_0$-junction
  behavior for a strongly spin-polarized ferromagnetic barrier. 
\end{abstract}

\maketitle

{\it Introduction.--}
The interplay between magnetism and superconductivity leads to a number of fascinating
phenomena.
However, it is nontrivial to observe since spin-singlet superconductivity is
typically destroyed by strong magnetic fields through orbital~\cite{Ginzburg56} or
Zeeman-induced pair breaking~\cite{Chandrasekhar, Clogston}. 
Recently, superconductivity was experimentally realized in various two-dimensional
transition-metal dichalcogenides~\cite{Lu15, Saito16, Xi16, Xing17, Dvir18,Costanzo18,
Lu18, delaBarrera18, Sohn18, Li20, Cho21, Hamill21, Idzuchi21, Ai21, Kang21, Kuz21}.
In these materials, the orbital depairing effect from an in-plane magnetic field is
suppressed due to their two-dimensional nature.
For an odd-number-layer crystal, inversion symmetry is broken so that the spin-orbit
interaction from the transition-metal atom leads to spin-valley locking, i.e., a
valley-dependent Zeeman-like spin splitting~\cite{MoS2_12}. 
Since the spins are polarized out of plane, this Zeeman-like field was termed Ising
spin-orbit coupling (ISOC). 
Its presence makes the superconducting state resilient against the Zeeman effect from an
in-plane magnetic field~\cite{Bulaevskii76, Gorkov01, Frigeri04} far beyond the Pauli
paramagnetic limit~\cite{Chandrasekhar, Clogston}. 
Thus, these so-called Ising superconductors provide an ideal laboratory to study the
interplay between superconductivity and ferromagnetism. Furthermore, applying an in-plane
magnetic field induces triplet correlations~\cite{Rahimi17, Moeckli18, Moeckli20,
Moeckli20_DOS, mirage, Kuz21}, mirage gaps~\cite{mirage}, or a two-fold rotational
symmetry of the superconducting state~\cite{Hamill21, Cho21}.

Very recently, van der Waals heterostructures consisting of Ising superconductors and
ferromagnetic barriers have attracted a great deal of experimental
interest~\cite{Idzuchi21, Ai21, Kang21}. In particular, ferromagnetic Josephson junctions
have been fabricated and the coexistence of $0$ and $\pi$ states in the junction region
has been demonstrated which can be used to construct $\phi$-phase Josephson
junctions~\cite{Idzuchi21, Ai21}. 
Transport properties of Ising superconductors have also been theoretically investigated in
ferromagnet--Ising-superconductor junctions~\cite{Zhou16, Transport_Sun18} and Josephson
junctions with a half-metal barrier~\cite{Transport_Sun19}.
These theoretical works focused on phenomena arising from spin-triplet Andreev reflection
at the interfaces. 
So far, however, the influence of spin-triplet pairing correlations induced by
ferromagnetism on the transport properties of Ising superconductors has not yet been
discussed. 

\begin{figure*}
\centering
\includegraphics[width=\textwidth]{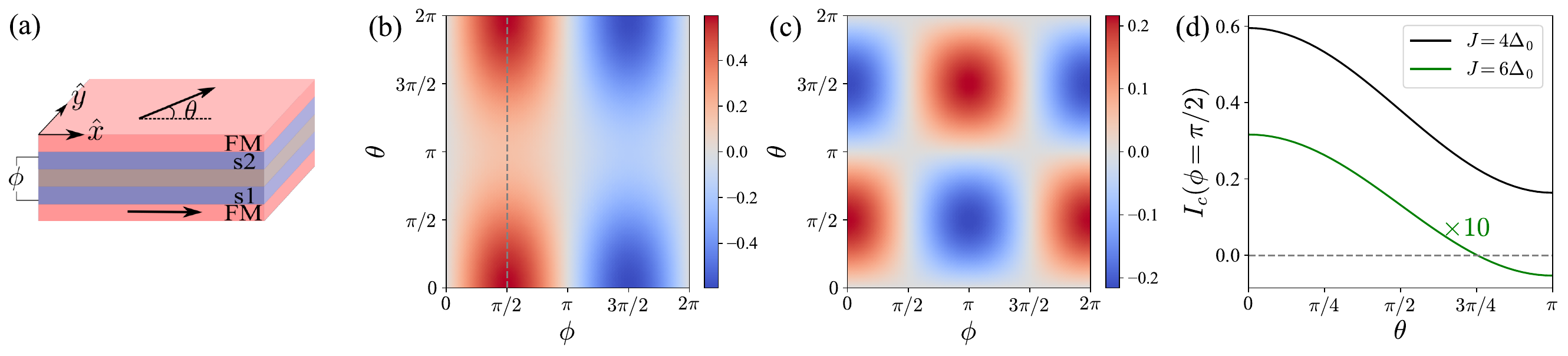}
\caption{(a) Schematic plot of the Ising-superconductor Josephson junction. 
  The ferromagnetic (FM) layers provide in-plane magnetic exchange fields in the two Ising
  superconductors which are labeled as $s1$ and $s2$. The angle between the exchange
  fields is $\theta$ and the superconducting phase difference is $\phi$. 
  The tunnel barrier separating the superconductors can be either nonmagnetic or
  ferromagnetic.
  (b) Charge current $I_c$ and (c) spin current $I_s$ as a function of $\phi$ and $\theta$
  for the junction with a nonmagnetic barrier.
  In (b) and (c), $\beta_{\rm so}=5\Delta_0$, $J=4\Delta_0$, and $T=0.01T_{0}$ where
  $\Delta_0$ and $T_{0}$ are, respectively, the zero-temperature gap and transition
  temperature in the absence of an external field.
  (d) Critical charge current versus angle $\theta$. 
  The black curve is the vertical line-cut along the dashed line of panel (b) at
  $\phi=\pi/2$. 
  The green curve with $J=6\Delta_0$ is multiplied by $10$ and shows a $0$-$\pi$
  transition at $\theta\approx 3\pi/4$. 
  The charge and spin currents are, respectively, in units of $G_t \Delta_0 / e$ and
  $\hbar G_t \Delta_0 /(2e^2)$ where $G_t$ is the tunnel conductance. }
\label{fig1}
\end{figure*}

In this Letter, we study the implications of in-plane exchange field-induced triplet
pairing correlations in a Josephson junction based on Ising superconductors [see
Fig.~\ref{fig1}(a)].
For a junction with a nonmagnetic barrier, we find that the charge supercurrent is
switchable by changing the exchange fields between parallel and antiparallel alignments.
At low temperatures and in the clean limit, a $\pi$-state charge supercurrent can be
realized if the exchange-field magnitudes are larger than the ISOC and superconductivity
is not fully destroyed. 
Noncollinear exchange fields give rise to a finite spin supercurrent. 
We also study the case of a ferromagnetic barrier and find that both the charge and spin
current-phase relations can be tuned if the barrier is strongly polarized.

{\it Model and formalism.--}
We consider an Ising superconductor with an $s$-wave paring gap $\Delta$ and
superconducting phase $\phi_\alpha$ in contact with a ferromagnetic layer. The effective
Bogoliubov-de Gennes Hamiltonian near one of the valleys can be written in the Nambu basis
$(c_{{\bm p},s,\uparrow}, c_{{\bm p},s,\downarrow}, c_{-{\bm p},-s,\uparrow}^\dag,
c_{-{\bm p},-s,\downarrow}^\dag)$ as
\begin{equation} \label{H}
  H({\bm p},s)  = 
  \begin{bmatrix}
    H_0({\bm p},s)  &  \Delta e^{i\phi_\alpha} i\sigma_y  \\
    -\Delta e^{-i\phi_\alpha} i\sigma_y  &  -H_0^*(-{\bm p},-s) 
  \end{bmatrix} ,
\end{equation}
where ${\bm p}$ is the momentum deviation from the $K$ or $K'$ point and $s=\pm$ denotes
the valley index. The normal-state Hamiltonian $H_0$ is 
\begin{equation}
  H_{0}({\bm p},s) = \xi_{\bm p}\sigma_0 + s\beta_{\rm so}\sigma_z 
  - {\bm J}\cdot {\bm \sigma} ,
\end{equation}
where the dispersion $\xi_{\bm p}=|{\bm p}|^2/(2m)-\mu$ is measured from the chemical
potential $\mu$ and $m$ is the electron mass. 
The Pauli matrices ${\bm \sigma}=(\sigma_x,\sigma_y,\sigma_z)$ act on spin space, with
$\sigma_0$ being the corresponding unit matrix.
By defining the out-of-plane direction along the $z$ axis, the ISOC is taken into account
via the term  $s\beta_{\rm so}\sigma_z$. 
The in-plane exchange field ${\bm J}=J_x {\bm x}+J_y {\bm y}$ from the
ferromagnetic layer gives rise to the Zeeman term $-{\bm J}\cdot {\bm \sigma}$ and
converts Cooper pairs from spin-singlet to spin-triplet~\cite{Moeckli20, mirage}.
In NbSe$_2$, the ISOC strength is about $40\,$meV for a monolayer and is decreasing
with the number of layers, while the Fermi energy is about $0.4\,$eV~\cite{Xi16,
Hamill21}.

For Ising superconductors, the superconducting gap and ISOC strength are much
smaller than the Fermi energy, and we can employ the quasiclassical
Green's function formalism~\cite{Eilenberger1968, LO1969, noneqSC, Belzig99} that provides
expressions of the charge and spin currents~\cite{circuit99, Belzig99, Eschrig15_NJP}. 
The structure of the quasiclassical Green's function, which is both valley and energy
$\varepsilon$ dependent, can be written as~\cite{noneqSC, Eschrig15}
\begin{equation} \label{g}
\hat{g}(s, \varepsilon) =
  \begin{bmatrix}
    g_0\sigma_0 + \bm{g}\cdot \bm{\sigma} & (f_0\sigma_0 + \bm{f}\cdot \bm{\sigma}) 
    i\sigma_y \\
    (\bar{f}_0\sigma_0 + \bar{\bm{f}} \cdot \bm{\sigma}^*)i\sigma_y  & \bar{g}_0
    \sigma_0+\bar{\bm{g}}\cdot {\bm \sigma}^* 
  \end{bmatrix} ,
\end{equation}
where the bar operation is defined as $\bar{q}(s, \varepsilon) = q(-s, -\varepsilon^*)^*$
with $q\in \{ g_0, f_0, {\bm g}, {\bm f}\}$. 
The anomalous Green's functions $f_0$ and $\bm{f}$ characterize the singlet and the
triplet pairings, respectively. 
As we show in the Supplemental Material~\cite{SM}, all components of $\hat{g}$ can be
obtained from the Eilenberger equation~\cite{Eilenberger1968, noneqSC, mirage} 
\begin{equation} \label{Eilenberger}
  \big[\varepsilon \tau_3 \sigma_0 -\hat{\Delta}-\hat{\nu} -\hat{\Sigma}(\varepsilon),
  \hat{g} \big] =0,
\end{equation}
together with ${\rm tr}(\hat{g})=0$ and the normalization condition
$\hat{g}^2=1$~\cite{mirage}. 
In Eq.~\eqref{Eilenberger}, $\tau_3$ is the third Pauli matrix acting on Nambu space, 
\begin{equation}
\hat{\Delta} = \Delta
  \begin{bmatrix}
   0 & e^{i\phi_\alpha} i\sigma_y \\ e^{-i\phi_\alpha} i\sigma_y & 0
  \end{bmatrix} ,
\end{equation}
and 
$\hat{\nu} = -{\rm diag}[{\bm J}\cdot {\bm \sigma}, {\bm J}\cdot {\bm \sigma}^*]
+ s\beta_{\rm so} \tau_3 \sigma_z$. 
Nonmagnetic intervalley impurity scattering is taken into account via
$\hat{\Sigma}(\varepsilon) = -i\Gamma\langle \hat{g}(s, \varepsilon) \rangle$ 
where $\Gamma$ is the impurity-scattering rate and $\langle \cdots \rangle$ denotes
averaging over all Fermi-momentum directions. 
In particular, we find ${\bm f}=a\ (\varepsilon {\bm J} +is\beta_{\rm so}{\bm z}\times{\bm
J})$ in the clean limit and the factor $a$ can be fixed by the normalization
condition~\cite{mirage, SM}. 
The second term in ${\bm f}$ originates from the commutator between the Zeeman term
${\bm J}\cdot {\bm \sigma}$ and the ISOC term $s\beta_{so}\sigma_z$ in the Green's
function.
The retarded counterpart of $\hat{g}(s, \varepsilon)$ is obtained by replacing
$\varepsilon$ with $\varepsilon +i\eta$, where $\eta$ is the Dynes broadening
parameter~\cite{Dynes}. 

We now turn to the Josephson junction which is schematically shown in Fig.~\ref{fig1}(a).
The in-plane exchange fields in the two Ising superconductors, which are denoted as $s1$
and $s2$, originate from the corresponding ferromagnetic layers. 
The two exchange fields, with a relative orientation defined by the angle $\theta$, are
assumed to have the same magnitude $J\equiv |{\bm J}|$. 
The magnetization direction of a ferromagnetic layer can be controlled by an
external magnetic field. For a junction in which the two ferromagnetic layers have
different thicknesses, an applied magnetic field will predominantly tune the magnetization
of the thinner layer leading to a controlled relative alignment between the exchange
fields~\cite{SC_spinvalve18}.
The Josephson phase is denoted by $\phi=\phi_{s1}-\phi_{s2}$. 
The central barrier separating the two superconductors can be either nonmagnetic or
ferromagnetic with out-of-plane magnetization and spin polarization ${\cal P}$. 
We consider a tunnel junction characterized by the conductance $G_t$ including the valley 
degree of freedom.
The expressions of the charge current $I_c$ and the $z$-polarized spin current $I_s$ in
superconductor $s1$ are, respectively, given by~\cite{Eschrig15_NJP, SM}
\begin{equation} \label{IcIs}
  I_c = \frac{G_t}{8e} \int_{-\infty}^{\infty} d\varepsilon \ {\rm tr}\big( \tau_3
  \sigma_0 \hat{I} \big),  \ \ 
  I_s = \frac{\hbar G_t}{16e^2} \int_{-\infty}^{\infty} d\varepsilon \ {\rm tr}\big(
  \tau_3 \sigma_z \hat{I} \big),
\end{equation}
where
\begin{align} \label{hat_I}
  \hat{I} = \frac{1}{2} & {\rm Re} \Big[\big(1+\sqrt{1-{\cal P}^2}\big)\hat{g}^r_{s2} +
  {\cal P} \{ \hat{\kappa}, \hat{g}^r_{s2}\} +  \notag \\
  & \big(1-\sqrt{1-{\cal P}^2}\big)\hat{\kappa}\hat{g}^r_{s2}\hat{\kappa}, \
  \hat{g}^r_{s1}\Big] \tanh[\varepsilon/(2k_BT)] , 
\end{align}
with $\hat{\kappa}={\rm diag}(\sigma_z, \sigma_z)$, $k_B$ the Boltzmann constant, and $T$
the temperature. 
Here, 
$\hat{g}^r_{\alpha}$ is the retarded counterpart of $\hat{g}$ in Eq.~\eqref{g} for
superconductor $\alpha=s1,s2$.  

\begin{figure*}
\centering
\includegraphics[width=5.2in]{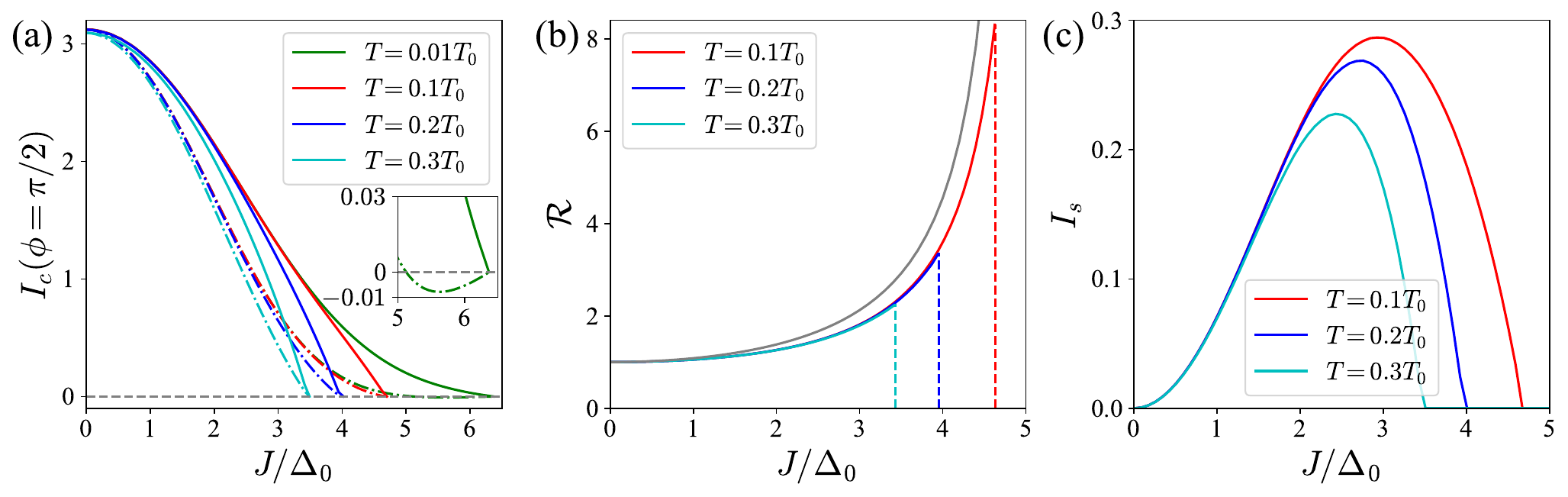}
\caption{(a) Critical charge current versus exchange-field magnitude $J$ at different
  temperatures. The solid and dash-dotted curves are for the configurations of parallel
  and antiparallel exchange fields, respectively. 
  The inset shows the curves at $T=0.01T_{0}$ for $J/\Delta_0 \geq 5$. The $0$-$\pi$
  transition occurs at $J\approx 5.1 \Delta_0$ for the antiparallel configuration. 
  (b) Switch ratio ${\cal R}$ versus exchange-field magnitude at different temperatures.
  The vertical dashed lines indicate the critical fields. 
  The gray line is the approximation of ${\cal R}$ given in Eq.~\eqref{R}. 
  (c) Spin current at $\phi=0$ and $\theta=\pi/2$ versus exchange-field magnitude. 
  Here, the barrier is nonmagnetic and $\beta_{\rm so}=5\Delta_0$. }
\label{fig2}
\end{figure*}

{\it Nonmagnetic barrier.--}
We first discuss the case of a tunnel junction with a nonmagnetic barrier. The charge
current $I_c$ and the spin current $I_s$ can be expressed as
\begin{align}
  I_c &= (I_{c0} + I_1 \cos\theta) \sin\phi , \label{Ic} \\
  I_s &= (\hbar/2e) (I_{s0} + I_1 \cos\phi) \sin\theta , \label{Is}
\end{align}
where $I_{c0}$, $I_{s0}$, and $I_1$ are derived in the Supplemental Material~\cite{SM}.
Equations~\eqref{Ic} and \eqref{Is} were previously obtained in Ref.~\cite{Linder15_PRB}
for an SNS-junction with a diffusive normal metal.
The charge and spin currents obey the relation $\partial I_c/\partial \theta =
(2e/\hbar)\partial I_s/\partial \phi$~\cite{Waintal02}. 
The terms proportional to $I_{c0}$ and $I_1$, respectively, result from singlet and
triplet pairing correlations. 
The phase-independent spin current component $(\hbar/2e)I_{s0}\sin\theta$ is due to the
noncollinear in-plane magnetizations induced by the exchange fields. 
Both $I_1$ and $I_{s0}$ vanish if one of the exchange fields is absent. 
Thus, the spin current can be modulated by changing the orientation $\theta$
or the magnitude $J$ of the in-plane exchange fields. Moreover, there is a pure spin
current at $\phi=0$ or $\phi=\pi$ for which the charge current vanishes. 
The charge current can be tuned by changing between the parallel and antiparallel
configurations of the exchange fields if the magnitude of $I_1$ is comparable to that of
$I_{c0}$. We will see that this is possible due to the presence of the ISOC fields in the
Ising superconductors. 

We assume that the ISOC fields have the same sign and the same magnitude at both sides of
the junction. The complementary case with opposite signs is considered in the Supplemental
Material~\cite{SM}. 
The superconducting gap is self-consistently calculated by neglecting the tunneling effect
between the two superconductors~\cite{mirage}.
The numerical results for the charge and spin currents are, respectively, shown in
Figs.~\ref{fig1}(b) and \ref{fig1}(c) in the clean limit. 
From Fig.~\ref{fig1}(b), we can clearly observe the difference of the critical charge
currents between the parallel and antiparallel configurations. 
This difference can be used to control the charge supercurrent, and we will call this
phenomenon the \textit{switch effect}.
The phase-independent part of the spin current $\hbar I_{s0}/(2e)$ is negligible compared
to the contribution proportional to $I_1$ as can be seen from Fig.~\ref{fig1}(c) at
$\phi=\theta=\pi/2$. 

For a clean Ising superconductor, there is an upturn for the critical magnetic field at
low temperatures~\cite{Saito16, Ilic17, Moeckli20, Liu20} so that the critical field can
be even larger than the ISOC. In this situation, the magnitude of $I_1$ can be larger than
that of $I_{c0}$ at $J \gtrsim \beta_{\rm so}$ so that the critical charge current can be
negative, and a $\pi$-state Josephson junction is realized. 
In Fig.~\ref{fig1}(d), the green curve shows the critical charge current versus the angle
$\theta$ at $J > \beta_{\rm so}$.
It can be seen that the critical charge current changes sign at $\theta \approx 3\pi/4$.
Thus, a $0$-$\pi$ transition can be achieved by changing the magnitudes or the relative
orientation of the exchange fields for weak impurity scattering and low temperature.
Typically, $\pi$-states are realized in Josephson junctions with a ferromagnetic barrier
between two superconductors~\cite{pi77, pi82, pi01, pi01_JETP, Buzdin05_RMP, Brydon09,
Alidoust10, Shomali_11, Gingrich2016, Transport_Sun19}. 
Here, its appearance is due to the interplay of the supercurrents induced by singlet and
triplet pairing correlations. 

In the following, we discuss the dependence of the switch effect of the charge current on
the temperature and exchange fields. 
Figure~\ref{fig2}(a) shows the critical charge currents versus the exchange-field
magnitude at different temperatures. 
The critical currents for the parallel and antiparallel exchange fields are shown in solid
and dash-dotted lines, respectively.
On increasing the exchange-field magnitude $J$, the critical charge currents for both the
parallel and antiparallel configurations decrease, which results from the suppression of
the superconducting gap $\Delta$. 
Meanwhile, the triplet pairing correlations first increase and then decrease due to the
interplay between the increase of the exchange field and the decrease of the
superconducting gap. This is reflected in the difference of the critical charge currents
between the parallel and antiparallel configurations with increasing $J$. 
The spin current at $\phi=0$ exhibits the same nonmonotonic behavior with respect to $J$
as shown in Fig.~\ref{fig2}(c).

To study the switch effect quantitatively, we define the ratio of the critical charge
currents between the parallel and antiparallel configurations, 
\begin{equation}
  {\cal R} = \frac{I_c(\theta=0)}{I_c(\theta=\pi)}\bigg|_{\phi=\pi/2}
  = \frac{I_{c0}+I_1}{I_{c0}-I_1} .
\end{equation}
The switch ratio versus the exchange-field magnitude is shown in Fig.~\ref{fig2}(b) where
the vertical lines indicate the critical fields at different temperatures. We see that the
switch ratio increases with increasing the exchange-field magnitude and can achieve large
values at low temperatures. The gray line shows the approximation~\cite{SM}
\begin{equation} \label{R}
  {\cal R} \approx (\beta_{\rm so}^2+J^2)/(\beta_{\rm so}^2-J^2) ,
\end{equation}
which neglects the pair-breaking effect. In this approximation, the switch ratio
diverges near $J=\beta_{\rm so}$ and can be negative for $J>\beta_{\rm so}$ indicating a
$\pi$-state for the antiparallel configuration.
This behavior can be seen in the inset of Fig.~\ref{fig2}(a) with the temperature being
close to zero. 

We have assumed that the ISOCs have the same sign on both sides of the junction. If the
signs are opposite, there is a quantitative change: the sign of $I_1$ is reversed so that
the critical charge current for the antiparallel configuration is larger than that for the
parallel configuration~\cite{SM}. This does not affect the existence of the switch effect
and the $0$-$\pi$ transition.

The switch effect in this work is due to the interplay between ISOC and exchange fields.
This differs from the effect described by Bergeret et al.~\cite{Bergeret01} in which the
critical current of the antiparallel configuration is larger than that of the parallel
configuration.

{\it Ferromagnetic barrier.--}
The switch ratio can be increased by reducing the supercurrent carried by the singlet
Cooper pairs. This can be achieved by replacing the nonmagnetic barrier with a
ferromagnetic one. 
We consider the case where the magnetization of the ferromagnetic barrier points out of
plane with spin polarization ${\cal P}$. 
The charge current $I_c$ and spin current $I_s$ are, respectively, expressed as~\cite{SM}
\begin{align}
  I_c &= \big(\sqrt{1-{\cal P}^2} I_{c0} + I_1 \cos\theta \big) \sin\phi 
  + {\cal P} I_1 \sin\theta \cos\phi , \label{Icc} \\
  I_s &= (\hbar/2e)\big[ (I_{s0} +  I_1 \cos\phi) \sin\theta
  + {\cal P} I_1 \sin\phi \cos\theta \big] , \label{Iss}
\end{align}
where $I_{c0}$, $I_{s0}$ and $I_1$ are the same coefficients as in Eqs.~\eqref{Ic} and
\eqref{Is} that are derived in the Supplemental Material~\cite{SM}. The charge current
carried by the singlet Cooper pairs becomes $\sqrt{1-{\cal P}^2}I_{c0}$ and is reduced
compared to that for a nonmagnetic barrier. 
This enhances the switch effect of the critical charge current between $\theta=0$ and
$\theta=\pi$. The $\pi$-state can also be realized even at $J<\beta_{\rm so}$. Moreover,
the ferromagnetic barrier leads to a phase shift to the supercurrent induced by the
triplet pairing correlations.

\begin{figure}
\centering
\includegraphics[width=\columnwidth]{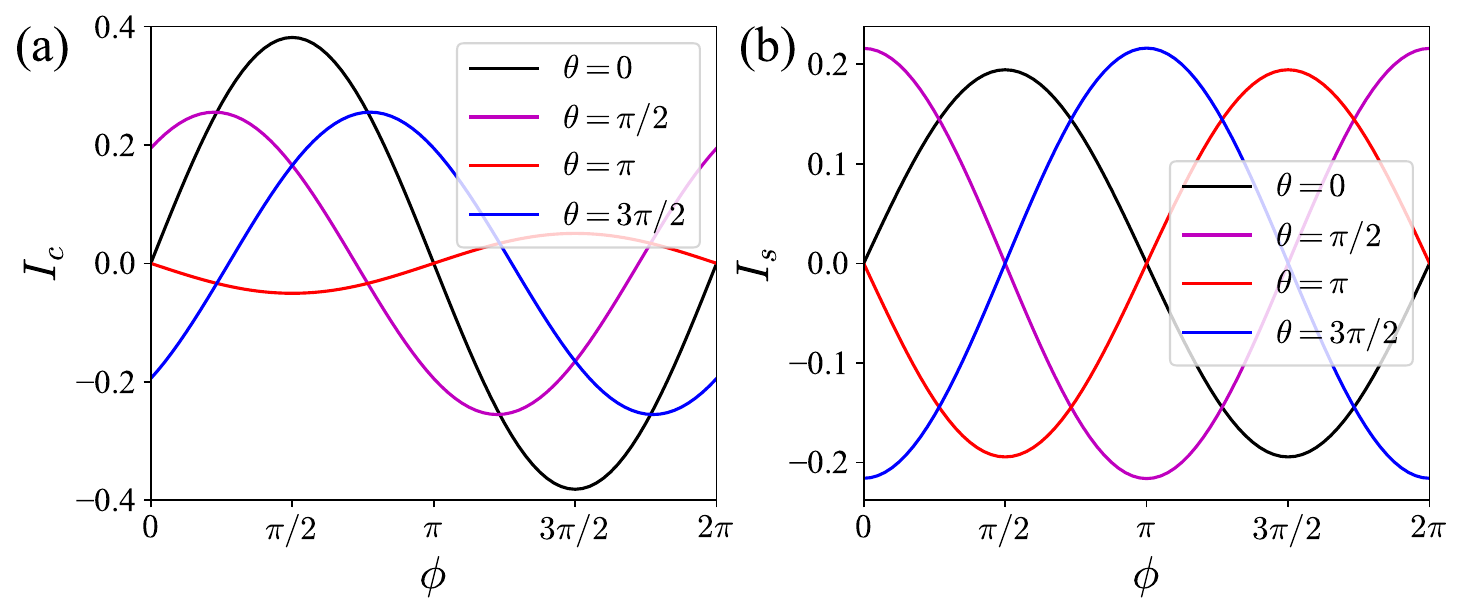}
\caption{(a) Charge current and (b) spin current for the junction with a ferromagnetic
  barrier of which the magnetization points out of plane and the spin polarization
  is ${\cal P}=0.9$. The other parameters are the same as those in Figs.~\ref{fig1}(b) and
  \ref{fig1}(c). }
\label{fig3}
\end{figure}

The charge and spin current-phase relations at different angles $\theta$ are shown in
Fig.~\ref{fig3}. We use the same parameters as in Figs.~\ref{fig1}(b) and \ref{fig1}(c)
apart from the finite spin polarization of the barrier. 
A $\pi$-state of the charge current arises for the antiparallel configuration
($\theta=\pi$) [see Fig.~\ref{fig3}(a)].
Moreover, as can be seen from Fig.~\ref{fig3}, both the charge and spin current-phase
relations can be controlled by the angle $\theta$ between the exchange fields. 
This tunability depends on the spin polarization of the ferromagnetic barrier.
When the ferromagnetic barrier is fully polarized (${\cal P}=\pm 1$), we get 
$I_c = I_1 \sin(\phi \pm \theta)$ and $I_s = (\hbar/2e)[I_{s0} \sin\theta + I_1\sin(\theta
\pm \phi)]$ indicating that the phases of both the charge and spin currents can be
arbitrarily tuned. Thus, the present system is a realization of a so-called
$\phi_0$-junction for charge and spin supercurrents~\cite{Eschrig15_NJP}. 

There are further interesting implications of Eqs.~\eqref{Icc} and \eqref{Iss}.
At Josephson phase $\phi=0$ or $\phi=\pi$, the supercurrent is only carried by the triplet
Cooper pairs and both the charge and spin currents depend on the angle between the
exchange fields in a sinusoidal way. 
Moreover, for the spin current, we have
\begin{align}
  & I_s(\theta=0) = -I_s(\theta=\pi) =(\hbar/2e){\cal P}I_1\sin\phi , \label{Is1} \\
  & I_s(\theta=\pi/2) = -I_s(\theta=3\pi/2) =(\hbar/2e)(I_{s0}+I_1\cos\phi) , \label{Is2}
\end{align}
which can be also seen from Fig.~\ref{fig3}(b). Equations~\eqref{Is1} and \eqref{Is2} show
that the spin current-phase relation can be switched between sine and cosine by changing
the relative orientation of the exchange fields.

{\it Discussion and conclusion.--}
For our investigation we only considered Ising superconductors in the clean limit.
Since the induced triplet pairing correlations are more sensitive to intervalley
scattering than the singlet pairing correlations~\cite{Moeckli20, mirage}, the switch
effect of the charge current is suppressed by impurity scattering~\cite{SM}.

In the presence of an in-plane exchange field, finite-energy pairing correlations emerge
in an Ising superconductor accompanied by mirage gaps~\cite{mirage}. In fact, for the
present discussion of the Josephson effect, the contribution from the mirage gaps to the
supercurrents is negligible compared to that from the main superconducting gap~\cite{SM}.

On the experimental side, Josephson junctions with ferromagnetic barriers based on
NbSe$_2$ with Ising superconductivity have been fabricated and studied~\cite{Idzuchi21,
Ai21, Kang21}.  Moreover, the magnetic proximity effect in van der Waals
heterostructures has been investigated. Thus, our predictions can be potentially checked
in experiments based on van der Waals heterostructures consisting of the Ising
superconductor NbSe$_2$ and a two-dimensional magnet such as CrBr$_3$~\cite{Hamill21}. 

To conclude, we have studied the transport properties of a van der Waals Josephson
junction consisting of Ising superconductors and ferromagnets.
The ferromagnetic layers provide in-plane magnetic exchange fields that induce
controllable triplet pairing correlations in the superconductors.
As a result, both the charge and spin currents can be modulated by the strength and
relative directions of the exchange fields. In particular, we have described a switch
effect for the charge current and a $0$-$\pi$ transition. 
Furthermore, both the charge and spin current-phase relations are tunable if the barrier
is strongly spin-polarized.
Our predictions show that Josephson junctions based on Ising superconductors exhibit rich
transport properties, and they confirm the great potential in van der Waals
superconducting heterostructures.

\bigskip

\begin{acknowledgments}
  We acknowledge useful discussions with A. Di Bernado and E. Scheer.
  G.T. and C.B. acknowledge financial support from the Swiss National Science Foundation
  (SNSF) and the NCCR Quantum Science and Technology.
  R.L.K. and W.B. acknowledge funding by the Deutsche Forschungsgemeinschaft (DFG, German
  Research Foundation) -- Project-ID 443404566 - SPP 2244. 
\end{acknowledgments}

\bibliography{bib_IsingSC}{}

\clearpage

\begin{center}
  \large{\bf{Supplemental Material for ``Controlling charge and spin transport in an
  Ising-superconductor Josephson junction"}}
\end{center}

\section{Quasiclassical Green's function}
We consider an Ising superconductor with a singlet $s$-wave pairing gap $\Delta$ and
superconducting phase $\phi_\alpha$. 
The effective Bogoliubov-de Gennes Hamiltonian near one of the valleys can be written in
the Nambu basis $(c_{{\bm p},s,\uparrow}, c_{{\bm p},s,\downarrow}, c_{-{\bm
p},-s,\uparrow}^\dag, c_{-{\bm p},-s,\downarrow}^\dag)$ as
\begin{equation} \label{H_BdG}
  H_{\mathrm{BdG}} = 
  \begin{bmatrix}
    H_0({\bm p},s)  &  \Delta e^{i\phi_\alpha} i\sigma_y  \\
    -\Delta e^{-i\phi_\alpha} i\sigma_y  &  -H_0^*(-{\bm p},-s) 
  \end{bmatrix} ,
\end{equation}
where ${\bm p}$ is the deviation from the $K$ or $K'$ point and $s=\pm$ denotes the valley
index. By defining the out-of-plane direction along the $z$ axis, the normal-state
Hamiltonian $H_0$ is 
\begin{equation}
  H_0({\bm p},s)=\xi_{\bm p}\sigma_0 +s\beta_{\rm so}\sigma_z -{\bm J}\cdot {\bm \sigma} ,
\end{equation}
where the dispersion $\xi_{\bm p}=|{\bm p}|^2/(2m)-\mu$ is measured form the chemical
potential $\mu$ and $m$ is the electron mass. The Pauli matrices ${\bm \sigma}=(\sigma_x,
\sigma_y, \sigma_z)$ act on spin space with $\sigma_0$ being the corresponding unit
matrix. 
The strength of the ISOC which pins the electron spins in the out-of-plane direction is
denoted by $\beta_{\mathrm{so}}$. The in-plane exchange field ${\bm J}= J_x {\bm x} + J_y
{\bm y}$ induces the Zeeman term $-{\bm J}\cdot {\bm \sigma}$. 

We employ the quasiclassical formalism which concentrates on the phenomena close to
the Fermi surface~\cite{Eilenberger1968, LO1969, Belzig99, noneqSC}. This is justified
since both the superconducting gap and the ISOC strength are much smaller than the Fermi
energy in a typical Ising superconductor. The exact quasiclassical Green's function for
the Ising superconductor with Zeeman term $-J_x\sigma_x$ was already derived in the
Supplemental Material of Ref.~\cite{mirage}. Here, we generalize this result to the
situation with an arbitrary in-plane direction. The structure of the quasiclassical
Green's function is written as~\cite{noneqSC, Eschrig15}
\begin{equation} \label{hat_g}
  \hat{g}(s, \varepsilon) =
  \begin{bmatrix}
    g_0\sigma_0 + \bm{g}\cdot \bm{\sigma} & (f_0\sigma_0 + \bm{f}\cdot \bm{\sigma}) 
    i\sigma_y \\
    (\bar{f}_0\sigma_0 + \bar{\bm{f}} \cdot \bm{\sigma}^*)i\sigma_y  & \bar{g}_0
    \sigma_0+\bar{\bm{g}}\cdot {\bm \sigma}^* 
  \end{bmatrix} ,
\end{equation}
where the energy $\varepsilon$ is relative to the Fermi energy and the bar operation is
defined as $\bar{q}(s, \varepsilon) = q(-s, -\varepsilon^*)^*$ with $q\in \{ g_0, f_0,
{\bm g}, {\bm f} \}$. The energy $\varepsilon$ is complex for the retarded and advanced
counterparts. The anomalous Green's functions $f_0$ and $\bm{f}$ characterize the singlet
and triplet pairings, respectively. 
Furhtermore, the condition ${\rm tr}\big( \hat{g} \big) =0$ leads to $\bar{g}_0 = -g_0$.
By introducing ${\bm g}_{\pm}=({\bm g}\pm \bar{\bm g})/2$, the normalization condition
$\hat{g} \hat{g} =\tau_0\sigma_0$ results in
\begin{align}
  &g_0^2 + \bm{g}_+^2 + {\bm g}_-^2 - f_0\bar{f}_0 +{\bm f}\cdot \bar{\bm f} =1,
  \label{norm} \\
  &2g_0 \bm{g}_+ = \bar{f}_0 {\bm f}- f_0 \bar{{\bm f}}, \label{g+} \\
  &2g_0 \bm{g}_- = i \bar{\bm f}\times {\bm f}. \label{g-}
\end{align}
Here and below, $\tau_1$, $\tau_2$, and $\tau_3$ are the Pauli matrices acting on Nambu
space with $\tau_0$ being the corresponding unit matrix. 
The term $\bm{g}_+$ which is valley independent describes the in-plane magnetization of
the Ising superconductor. As we will see later, $\bm{g}_-$ only contains $g_{-,z}$
component that characterizes the spin-polarization between the $K$ and $K'$ valleys in $z$
direction. 

For a homogeneous system in the clean limit, the quasiclassical Green's function
$\hat{g}(s, \varepsilon)$ obeys the Eilenberger equation~\cite{Eilenberger1968, noneqSC}
\begin{equation} \label{Eilen_clean}
  \big[\varepsilon \tau_3\sigma_0 -\hat{\Delta}-\hat{\nu}, \hat{g} \big] =0. 
\end{equation}
Here, the order-parameter term $\hat{\Delta}$ is explicitly written as 
\begin{equation}
\hat{\Delta} = 
  \begin{bmatrix}
   0 & \Delta e^{i\phi_\alpha} i\sigma_y \\ \Delta e^{-i\phi_\alpha} i\sigma_y & 0
  \end{bmatrix} .
\end{equation}
The in-plane exchange field and the ISOC are included in $\hat{\nu}$ as
\begin{equation}
  \hat{\nu} =
  \begin{bmatrix}
    ({\bm \nu}_+ + {\bm \nu}_-)\cdot {\bm \sigma} & 0 \\
     0 & ({\bm \nu}_+ - {\bm \nu}_-)\cdot {\bm \sigma}^*  
  \end{bmatrix} 
\end{equation}
with
${\bm \nu}_+ = -{\bm J}$ and ${\bm \nu}_- = (0, 0, s\beta_{\rm so})$.
The off-diagonal terms of the Eilenberger equation in Nambu space are
\begin{align}
  &\varepsilon f_0 -{\bm \nu}_+\cdot {\bm f} +\Delta g_0 =0, \label{f0} \\
  &\varepsilon {\bm f} - {\bm \nu}_+ f_0-i{\bm \nu}_-\times {\bm f} +\Delta {\bm
  g}_+ =0, \label{ff} \\
  &\varepsilon \bar{f}_0 +{\bm \nu}_+\cdot \bar{\bm f} +\Delta g_0 =0,
  \label{bf0} \\
  &\varepsilon \bar{\bm f} +{\bm \nu}_+\bar{f}_0 +i{\bm \nu}_-\times \bar{\bm f} 
  -\Delta {\bm g}_+ =0, \label{bff}
\end{align}
where the phase factors of $(f_0, {\bm f})$ and $(\bar{f}_0, \bar{\bm f})$
are dropped out for simplicity. 
Equations~\eqref{f0} and \eqref{ff} can be written explicitly as
\begin{equation} \label{f0ff}
  \begin{bmatrix}
    \varepsilon & J_x & J_y \\
    J_x & \varepsilon & is\beta_{\rm so} \\
    J_y & -is\beta_{\rm so} & \varepsilon \\
  \end{bmatrix}
  \begin{bmatrix}
    f_0 \\ f_x \\ f_y 
  \end{bmatrix} 
  + \Delta 
  \begin{bmatrix}
    g_0 \\ g_{+,x} \\ g_{+,y} \\
  \end{bmatrix} 
  =0 .
\end{equation}
The fact that $f_z=0$ leads to $g_{-,x}=g_{-,y}=g_{+,z}=0$ from Eqs.~\eqref{g+} and
\eqref{g-}. 

By neglecting the superconducting phase factors, we obtain $f_0 = \bar{f}_0$. 
Combining this with Eqs.~\eqref{f0} and \eqref{bf0}, we have
\begin{equation}
  J_x (f_x + \bar{f}_x) + J_y (f_y + \bar{f}_y) = 0 .
\end{equation}
By combining Eqs.~\eqref{ff} and \eqref{bff}, we have
\begin{align}
  & \varepsilon \big[ J_y(f_x + \bar{f}_x) - J_x(f_y + \bar{f}_y) \big] \notag \\
  =& is\beta_{\rm so} \big[ J_y(f_y - \bar{f}_y) + J_x(f_x - \bar{f}_x) \big] .
\end{align}
Using the requirement $\bar{\bm f}(s, \varepsilon) = {\bm f}(-s, -\varepsilon^*)^*$,
one can get
\begin{align} 
  f_x &= a\ (\varepsilon J_x - is\beta_{\rm so} J_y) , \quad \bar{f}_x = -f_x^* , \\
  f_y &= a\ (\varepsilon J_y + is\beta_{\rm so} J_x) , \quad \bar{f}_y = -f_y^* . 
\end{align}
where $a$ is to be fixed by the normalization condition. 
The above relations can alternatively be expressed as
\begin{equation} 
  {\bm f} = a\ (\varepsilon {\bm J} - is\beta_{\rm so} {\bm J}\times {\bm z}) , \qquad
  \bar{\bm f} = -{\bm f}^*, 
\end{equation}
as provided in the main text.
Consequently, Eq.~\eqref{f0ff} is rewritten as
\begin{align}
  & \varepsilon f_0 + a\ \varepsilon J^2 + \Delta g_0 = 0, \label{E1} \\
  & J_x f_0 + a\ J_x (\varepsilon^2 -\beta_{\rm so}^2) +\Delta g_{+,x} =0, \label{E2} \\
  & J_y f_0 + a\ J_y (\varepsilon^2 -\beta_{\rm so}^2) +\Delta g_{+,y} =0. \label{E3}
\end{align}
Combining Eqs.~\eqref{E1}-\eqref{E3} with Eq.~\eqref{g+}, which can be written as
\begin{equation}
  g_0 g_{+,x} = a\ \varepsilon J_x f_0 , \quad  g_0 g_{+,y} = a\ \varepsilon J_y f_0 , 
\end{equation}
we obtain $g_0$ and $f_0$ as
\begin{equation} \label{g0f0}
  g_0 = a\ \varepsilon \ c/(2\Delta), \quad 
  f_0 = -a\ (J^2 + c/2), 
\end{equation}
where
\begin{equation}
  c = \varepsilon^2-\beta_{\rm so}^2-J^2-\Delta^2 + u ,
\end{equation}
with 
\begin{equation} \label{u}
  u =  \sqrt{(\varepsilon^2-\beta_{\rm so}^2-J^2-\Delta^2)^2-4J^2\Delta^2} .
\end{equation}
The term $g_{-,z}$ is obtained from Eq.~\eqref{g-} with
\begin{equation}
  g_0 g_{-,z} = a^2\ \varepsilon J^2 \ s\beta_{\rm so} . 
\end{equation}
The coefficient $a$ is fixed by Eq.~\eqref{norm}, which is 
\begin{equation}
  g_0^2 +g_{+,x}^2 +g_{+,y}^2 +g_{-,z}^2 -f_0\bar{f}_0 +f_x\bar{f}_x +f_y\bar{f}_y =1 ,
\end{equation}
so that
\begin{equation}
  a^2 (4 J^2\Delta^2 -c^2)\big[ \Delta^2 (2 J^2 +c)^2 -c^2\varepsilon^2 +4 J^2
  \beta_{\rm so}^2 \Delta^2 \big] = 4 c^2 \Delta^2 ,
\end{equation}
which can be simplified as
\begin{equation}
  a^2 =\frac{c \Delta^2}{u^2 [ c(\varepsilon^2-\Delta^2)-2\Delta^2J^2 ]}.
\end{equation}
We should be reminded that $(f_0, {\bm f})$ and $(\bar{f}_0, \bar{\bm f})$ contain the
phase factors $e^{i\phi_\alpha}$ and $e^{-i\phi_\alpha}$, respectively, which are
neglected in the derivation above for simplicity. The phase factors are important when
considering a Josephson junction. 
To this end, we obtain all the components of the quasiclassical Green's function
$\hat{g}$. The retarded and advanced counterparts of $\hat{g}(s,\varepsilon)$ are
obtained, respectively, by replacing the real $\varepsilon$ in $\hat{g}(s,\varepsilon)$
with $\varepsilon + i\eta$ and $\varepsilon - i\eta$, where $\eta$ is an infinitesimal
positive number.

\section{Josephson junction with nonmagnetic barrier}
We now consider a Josephson junction consisting of two Ising superconductors separated by
a nonmagnetic barrier. 
The corresponding ferromagnetic layer induces an in-plane exchange field in each Ising
superconductor through magnetic proximity effect.  
We assume that the exchange fields applied to the superconductors have the same magnitude.
The superconducting phase difference is denoted as $\phi=\phi_{s1}-\phi_{s2}$ and the
relative angle between the two in-plane exchange fields as $\theta$.
The transparency of the junction is denoted as $D$. 
From the quantum circuit theory~\cite{circuit99}, the expressions of the charge current
$I_c$ and the $z$-polarized spin current $I_s$ are, respectively, given by 
\begin{align}
  I_c &= \frac{G_0}{8e}\int_{-\infty}^{\infty} d\varepsilon \ 
  {\rm tr}\big(\tau_3\sigma_0\hat{I}^k \big), \\
  I_s &= \frac{\hbar G_0}{16e^2}\int_{-\infty}^{\infty} d\varepsilon \ 
  {\rm tr}\big(\tau_3\sigma_z\hat{I}^k \big), \label{I_z}
\end{align}
with $G_0=2e^2/h$ the conductance quantum. 
In above, $\hat{I}^k$ is the Keldysh component of the matrix current given by
\begin{equation} \label{Ik}
  \hat{I}^k = 2 \big( \hat{A}^r \hat{X}^k + \hat{A}^k \hat{X}^a \big)
\end{equation}
with 
\begin{equation}
  \check{A} = 2 D \big[\check{g}_{s2}, \check{g}_{s1}\big] , \quad 
  \check{X} = \Big[4 - D\big(2-\big\{\check{g}_{s2}, \check{g}_{s1}\big\}\big)\Big]^{-1} .
\end{equation}
We have included an additional factor of $2$ in Eq.~\eqref{Ik} to take the valley degree
of freedom into account. 
Here, $\check{g}_\alpha$ with $\alpha = s1, s2$ is in Keldysh space and has the
structure 
\begin{equation}
  \check{g}_\alpha = 
  \begin{pmatrix}
    \hat{g}^r_\alpha & \hat{g}^k_\alpha \\ 0 & \hat{g}^a_\alpha
  \end{pmatrix} ,
\end{equation}
where $\hat{g}^r_\alpha$ and $\hat{g}^a_\alpha$ are, respectively, the retarded and
advanced counterparts of $\hat{g}_\alpha(s,\varepsilon)$. The Keldysh component
$\hat{g}^k_\alpha$ is obtained via the relation $\hat{g}^k_\alpha = (\hat{g}^r_{\alpha} -
\hat{g}^a_{\alpha})\tanh[\varepsilon/(2k_BT)]$, where $k_B$ is the Boltzmann constant and
$T$ is the temperature. 

In the tunneling limit, i.e., $D\ll 1$, one has
\begin{equation}
  \hat{I}^k = 2D \ {\rm Re}\big(\big[ \hat{g}^r_{s2}, \hat{g}^r_{s1} \big]\big)
  \tanh[\varepsilon/(2k_BT)] ,
\end{equation}
and the expressions of the charge current $I_c$ and the $z$-polarized spin current $I_s$
are, respectively, given by
\begin{equation} \label{Ics}
  I_c = \frac{G_t}{8e} \int_{-\infty}^{\infty} d\varepsilon \ 
  {\rm tr}\big( \tau_3 \sigma_0 \hat{I} \big), \ \
  I_s = \frac{\hbar G_t}{16e^2} \int_{-\infty}^{\infty} d\varepsilon \ 
  {\rm tr}\big( \tau_3 \sigma_z \hat{I} \big),
\end{equation}
with the tunnel conductance $G_t = 2 D G_0$ and 
\begin{equation} \label{hat_I_SM}
  \hat{I} = {\rm Re}\big(\big[ \hat{g}^r_{s2}, \hat{g}^r_{s1} \big]\big)
  \tanh[\varepsilon/(2k_BT)] . 
\end{equation}
By expressing the retarded Green's function as
\begin{equation}
  \hat{g}^r_{\alpha}(s, \varepsilon) =
  \begin{bmatrix}
    g_{0,\alpha}\sigma_0 + \bm{g}_{\alpha}\cdot \bm{\sigma} & 
    (f_{0,\alpha}\sigma_0 + \bm{f}_{\alpha}\cdot \bm{\sigma}) i\sigma_y \\
    (\bar{f}_{0,\alpha}\sigma_0 + \bar{\bm f}_{\alpha} \cdot \bm{\sigma}^*)i\sigma_y &
    \bar{g}_{0,\alpha} \sigma_0+\bar{\bm g}_{\alpha}\cdot {\bm \sigma}^* 
  \end{bmatrix} ,
\end{equation}
where the superscripts ``$r$" in all the components have been dropped out, we have
\begin{align}
  & {\rm tr}\big(\tau_3 \sigma_0\big[\hat{g}^r_{s2}, \hat{g}^r_{s1}\big]\big) \notag \\
  =& 4 \big( f_{0,s1} \bar{f}_{0,s2} - f_{0,s2} \bar{f}_{0,s1}
  - \bm{f}_{s1} \cdot \bar{\bm{f}}_{s2} + \bm{f}_{s2} \cdot \bar{\bm{f}}_{s1} \big),
\end{align}
and
\begin{align}
  & {\rm tr}\big(\tau_3 \sigma_z\big[\hat{g}^r_{s2}, \hat{g}^r_{s1}\big]\big) \notag \\
  =& 4i \big( 2\bm{g}_{+,s2} \times {\bm g}_{+,s1} +
  \bm{f}_{s2} \times \bar{\bm f}_{s1} - {\bm f}_{s1} \times \bar{\bm f}_{s2} \big)_z .
\end{align}
By considering the case where the ISOC strengths of both superconductors are the same at
the same valley, we have in the clean limit the expressions:
\begin{align} 
  & {\rm Re}\big[{\rm tr}\big(\tau_3 \sigma_0\big[\hat{g}^r_{s2},
  \hat{g}^r_{s1}\big]\big)\big] \notag \\
  =& -8\, {\rm Im}\big[ f_0\bar{f}_0 + a^2 (\varepsilon^2 + \beta_{\rm so}^2)
  J^2 \cos\theta \big] \sin\phi 
\end{align}
and 
\begin{align}
  & {\rm Re}\big[{\rm tr}\big(\tau_3\sigma_z \big[\hat{g}^r_{s2},
  \hat{g}^r_{s1}\big]\big)\big] \notag \\
  =& -8\, {\rm Im}\big[ -{\bm g}_{+}^2 + a^2 (\varepsilon^2 + \beta_{\rm so}^2) J^2
  \cos\phi \big] \sin\theta.
\end{align}
Here, we have used the fact ${\bm g}_{+,s1}^2={\bm g}_{+,s2}^2\equiv {\bm g}_{+}^2$. 
Then we arrive at 
\begin{align}
  I_c &= (I_{c0} + I_1 \cos\theta) \sin\phi , \label{I_c} \\
  I_s &= (\hbar/2e)(I_{s0} +  I_1 \cos\phi) \sin\theta \label{I_s} ,
\end{align}
as presented in Eqs.~(8) and (9) in the main text, where the critical currents read  
\begin{align}
  I_{c0} &= -\frac{G_t}{e} \int_{-\infty}^{\infty} d\varepsilon \ 
  {\rm Im}\big( f_0\bar{f}_0 \big) \tanh[\varepsilon/(2k_BT)] , \\
  I_{s0} &= \frac{G_t}{e} \int_{-\infty}^{\infty} d\varepsilon \ 
  {\rm Im}\big( {\bm g}_+^2 \big) \tanh[\varepsilon/(2k_BT)] , \\
  I_1 &= -\frac{G_t}{e} \int_{-\infty}^{\infty} d\varepsilon \ 
  b (\varepsilon^2 + \beta_{\rm so}^2) \tanh[\varepsilon/(2k_BT)] ,
\end{align}
with $b = {\rm Im}(a^2) J^2$. 
It can be seen that the term of $I_1$ originates from the triplet correlation functions.
The expressions of $x$ and $y$-polarized spin currents can be obtained by replacing
$\tau_3 \sigma_x$ in Eq.~\eqref{I_z} with $\tau_3 \sigma_x$ and $\tau_0 \sigma_y$,
respectively. One can find that both the $x$ and $y$-polarized spin currents vanish by
taking the contributions from the two different valleys into account. 

In the tunneling limit, we define the ratio of the charge currents between the parallel
and the antiparallel configurations as
\begin{equation}
  {\cal R} = \frac{I_c(\theta=0)}{I_c(\theta=\pi)}\bigg|_{\phi=\pi/2} 
  = \frac{I_{c0}+I_1}{I_{c0}-I_1} .
\end{equation}
From Eq.~\eqref{I_c}, one has
\begin{equation}
  {\cal R} = \frac{\int_{-\infty}^{\infty} d\varepsilon \ {\rm Im}\big[ f_0\bar{f}_0 + a^2
  J^2 (\varepsilon^2 + \beta_{\rm so}^2)\big]
  \tanh[\varepsilon/(2k_BT)]}{\int_{-\infty}^{\infty} d\varepsilon \ {\rm Im}\big[
  f_0\bar{f}_0 -a^2 J^2 (\varepsilon^2+\beta_{\rm so}^2)\big] \tanh[\varepsilon/(2k_BT)]}.
\end{equation}
Since $\Delta$ is small compared to $J$ and $\beta_{\rm so}$, Eq.~\eqref{u} can be
approximated as $u \approx \varepsilon^2-\beta_{\rm so}^2-J^2-\Delta^2$ so that the term
$f_0$ in Eq.~\eqref{g0f0} becomes
\begin{equation}
  f_0 \approx -a(\varepsilon^2 -\beta_{\rm so}^2 -\Delta^2) .
\end{equation}
In the tunneling limit, the Andreev bound states are localized around $\varepsilon=\pm
\Delta$ so that
\begin{equation}
  {\cal R} \approx \frac{ \beta_{\rm so}^4 + J^2 (\Delta^2 + \beta_{\rm so}^2)}{
  \beta_{\rm so}^4 - J^2 (\Delta^2+\beta_{\rm so}^2)} ,
\end{equation}
where we have ignored the contribution from the mirage gaps~\cite{mirage}. With $\Delta\ll
\beta_{\rm so}$, we arrive at
\begin{equation} \label{RR}
  {\cal R} \approx (\beta_{\rm so}^2+J^2)/(\beta_{\rm so}^2-J^2),
\end{equation}
as presented in Eq.~(11) in the main text.



\begin{figure}
\centering
\includegraphics[width=\columnwidth]{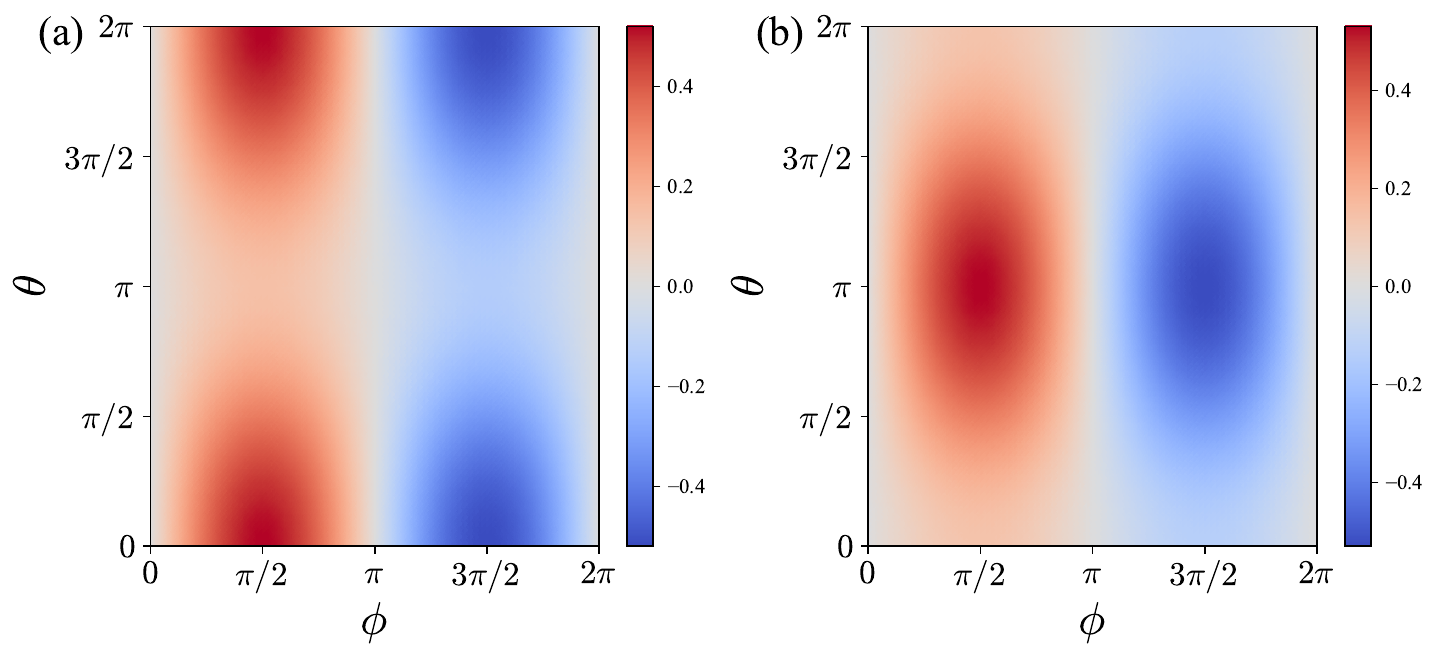}
\caption{Charge currents, in units of $G_t \Delta_0 / e$, are shown against the phase
  difference $\phi$ and the relative exchange-field angle $\theta$ in the tunneling limit. 
  The ISOCs have the same sign on both sides of the junction in (a) while the signs of the
  ISOCs are opposite in (b). 
  Here, $|\beta_{\rm so}|=5\Delta_0$, $J=4\Delta_0$ and $T=0.1T_{0}$ where $\Delta_0$ and
  $T_{0}$ are, respectively, the zero-temperature gap and transition temperature in the
  absence of an exchange field. }
\label{figS2}
\end{figure}

When the ISOC fields of two superconductors are opposite in sign at the same valley, $I_1$
in Eqs.~\eqref{I_c} and \eqref{I_s} become
\begin{equation}
  I_1 = -\frac{G_t}{e} \int_{-\infty}^{\infty} d\varepsilon \ 
  b (\varepsilon^2 - \beta_{\rm so}^2) \tanh[\varepsilon/(2k_BT)] .
\end{equation}
The charge currents for this case are shown in Fig.~\ref{figS2}(b). 
For comparison, we also show the case that the ISOC fields have the same sign at both
sides of the junction in Fig.~\ref{figS2}(a). 
Being different from Fig.~\ref{figS2}(a), the critical charge current in
Fig.~\ref{figS2}(b) is maximal (minimal) for the configuration of antiparallel (parallel)
exchange fields. 

\section{The contribution from mirage gaps}
\begin{figure}
\centering
\includegraphics[width=\columnwidth]{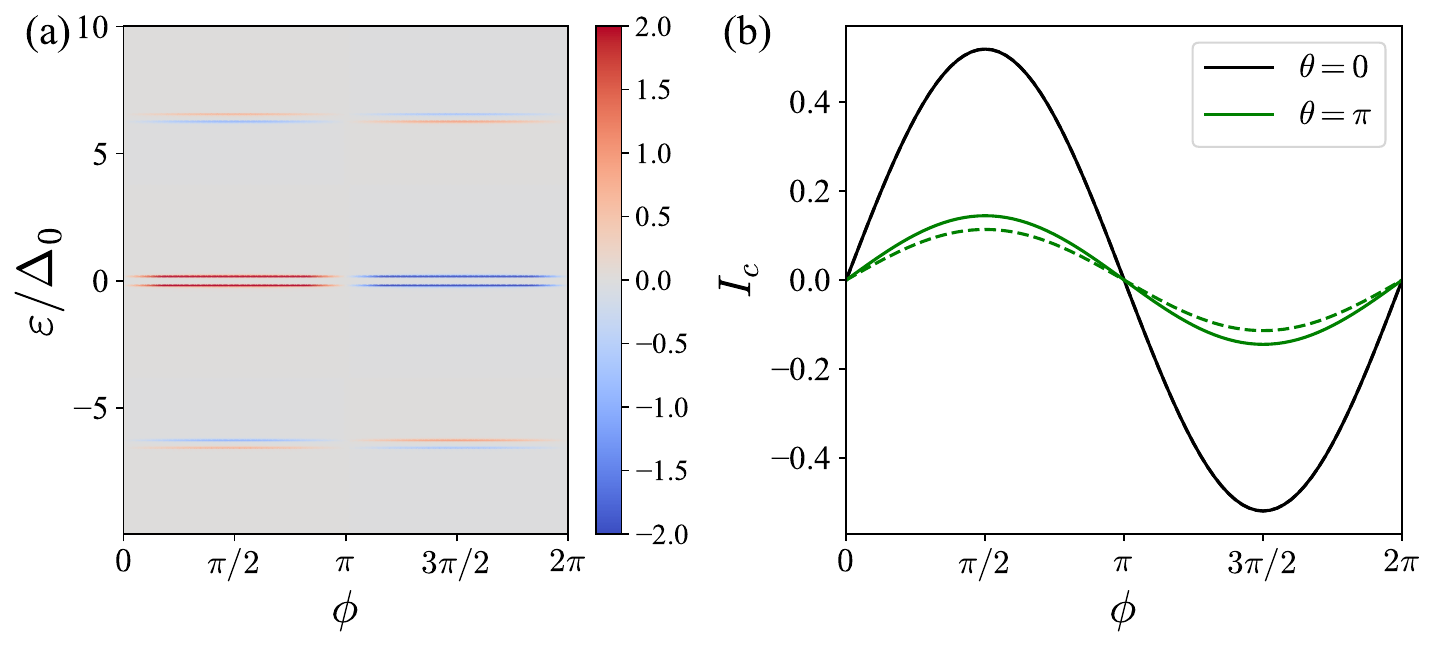}
\caption{(a) Spectral charge current $j_c(\varepsilon)$, in units of $G_t/e$,
  against energy $\varepsilon$ and Josephson phase $\phi$ at $\theta=0$.
  (b) Comparison of the charge currents between the cases with (solid lines) and without
  (dashed lines) the mirage gaps being considered. 
  The solid and dashed lines coincide with each other at $\theta=0$.  
  For the case without considering the contribution from the mirage gaps, the integration
  region to get the charge current is $\varepsilon \in [-4\Delta_0, 4\Delta_0]$. 
  Here, $\beta_{\rm so}=5\Delta_0$, $J=4\Delta_0$, and $T=0.1T_{0}$. }
\label{figS3}
\end{figure}

We define the spectral charge current $j_c(\varepsilon)$ through 
$I_c = \int_{-\infty}^{\infty} d\varepsilon j_c(\varepsilon)$.
In Fig.~\ref{figS3}(a), the spectral charge current $j_c(\varepsilon)$ is shown against
energy $\varepsilon$ and Josephson phase $\phi$ at $\theta=0$. 
We can see the appearance of Andreev bound states at energies larger than the
superconducting gap. 
This is due to the presence of the mirage gaps where the finite-energy pairings
occur~\cite{mirage}. 
In Fig.~\ref{figS3}(b), we compare the charge currents between the scenarios with and
without the mirage gaps being considered. It can be seen that the influence of the mirage
gaps on the charge current is small. 

\section{Disorder effect}
In the presence of nonmagnetic impurities, Eq.~\eqref{Eilen_clean} becomes
\begin{equation} 
  \big[\varepsilon \tau_3 \sigma_0 -\hat{\Delta}-\hat{\nu} -\hat{\Sigma}(\varepsilon),
  \hat{g} \big] =0,
\end{equation}
with $\hat{\Sigma}(\varepsilon) = -i\Gamma\langle \hat{g}(s, \varepsilon) \rangle$, 
where $\Gamma$ is the impurity-scattering rate and $\langle \cdots \rangle$ denotes
averaging over all Fermi-momentum directions. 

\begin{figure}
\centering
\includegraphics[width=\columnwidth]{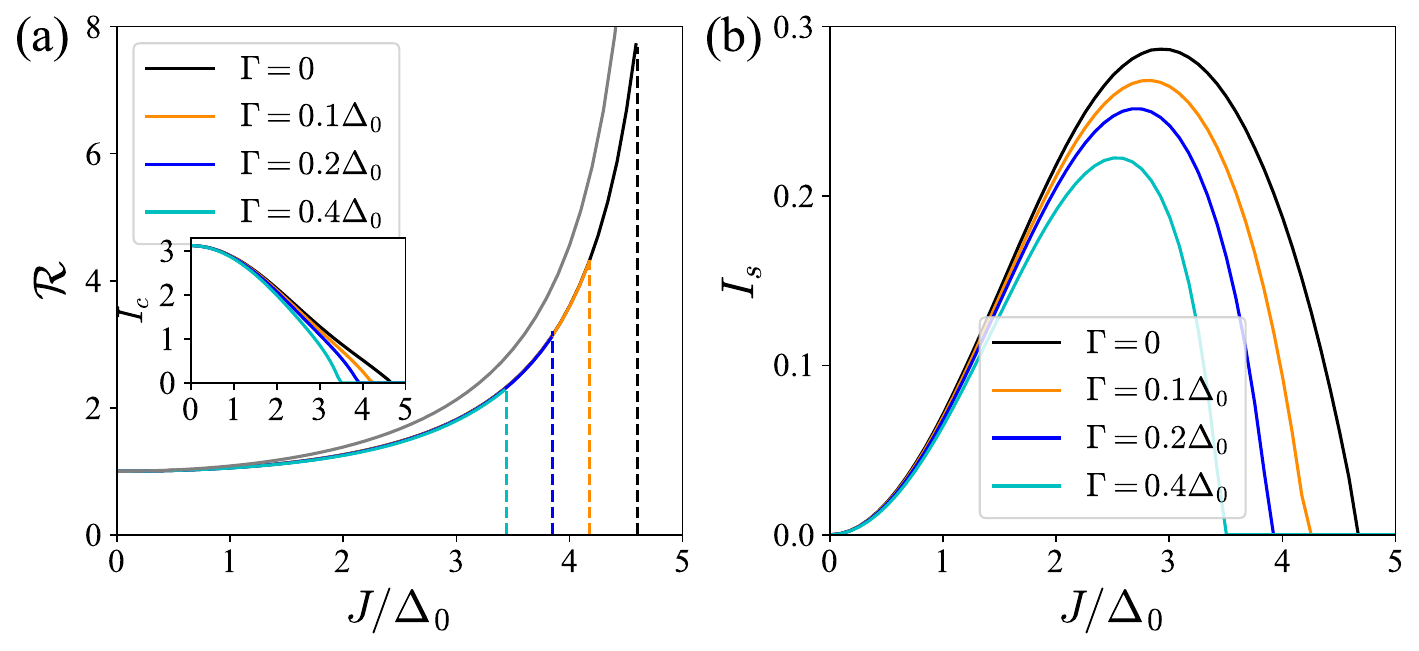}
\caption{(a) Switch ratio ${\cal R}$ versus exchange-field magnitude at different
  intervalley scattering rates $\Gamma$ in the tunneling limit. 
  The gray line is the approximation in Eq.~\eqref{RR}. 
  The inset shows the charge currents for the parallel configuration. 
  (d) Spin current at $\phi=0$ versus exchange-field magnitude. 
  Here, $\beta_{\rm so}=5\Delta_0$ and $T=0.1T_{0}$. }
\label{figS4}
\end{figure}

We consider the effects of nonmagnetic intervalley scattering in Fig.~\ref{figS4}. 
It can be seen that the intervalley scattering reduces the maximal switch ratio and also
the spin-current magnitude. 

\section{Josephson junction with ferromagnetic barrier}
We turn to the scenario with a ferromagnetic barrier which is characterized by the
magnetization vector ${\bm m}$ with $|{\bm m}|=1$ and the spin polarization ${\cal P}$. 
In the tunneling limit, $\hat{I}$ in Eq.~\eqref{Ics} is~\cite{Eschrig15_NJP}
\begin{align}
  \hat{I} = \frac{1}{2} & {\rm Re} \Big[\big(1+\sqrt{1-{\cal P}^2}\big)\hat{g}^r_{s2} +
  {\cal P} \{ \hat{\kappa}, \hat{g}^r_{s2}\} +  \notag \\
  & \big(1-\sqrt{1-{\cal P}^2}\big)\hat{\kappa}\hat{g}^r_{s2}\hat{\kappa}, \
  \hat{g}^r_{s1}\Big] \tanh[\varepsilon/(2k_BT)] , 
\end{align}
with the spin matrix
\begin{equation}
  \hat{\kappa} ={\rm diag} ({\bm m}\cdot {\bm \sigma}, \ \ {\bm m}\cdot {\bm \sigma}^*).
\end{equation}
Using the relations
\begin{equation}
  \{\hat{\kappa}, \hat{g}^r\} = 2
  \begin{bmatrix}
    (\bm{m}\cdot{\bm g})\sigma_0 + g_0 {\bm m}\cdot{\bm \sigma} & 
    i(\bm{m}\times\bm{f})\cdot\bm{\sigma} i\sigma_y \\
    -i(\bm{m}\times\bar{\bm{f}})\cdot\bm{\sigma}^* i\sigma_y & 
    (\bm{m}\cdot\bar{\bm g})\sigma_0 + \bar{g}_0 {\bm m}\cdot{\bm \sigma}^*
  \end{bmatrix} 
\end{equation}
and 
\begin{widetext}
\begin{equation}
  \hat{\kappa}\hat{g}^r\hat{\kappa} =
  \begin{bmatrix}
    g_0\sigma_0 + {\bm g}\cdot {\bm \sigma} &
    -[f_0\sigma_0 + 2(\bm{m}\cdot\bm{f})\bm{m}\cdot\bm{\sigma} -\bm{f}\cdot\bm{\sigma}]
    i\sigma_y \\
    -[\bar{f}_0\sigma_0 + 2(\bm{m}\cdot\bar{\bm{f}})\bm{m}\cdot\bm{\sigma}^*
    -\bar{\bm{f}}\cdot\bm{\sigma}^* ] i\sigma_y & 
    \bar{g}_0\sigma_0 + \bar{\bm g}\cdot {\bm \sigma}^*
  \end{bmatrix} ,
\end{equation}
\end{widetext}
separately, we obtain
\begin{equation}
  {\rm tr}\big(\tau_3\sigma_0\big[ \{\hat{\kappa}, \hat{g}_{s2}^r\}, 
  \hat{g}_{s1}^r\big]\big) = 8i \big[ (\bm{m}\times\bm{f}_{s2})\cdot\bar{\bm{f}}_{s1} -
  (\bm{m}\times\bm{f}_{s1})\cdot\bar{\bm{f}}_{s2} \big]
\end{equation}
and 
\begin{align}
  & {\rm tr}\big( \tau_3\sigma_0\big[ \hat{\kappa}\hat{g}_{s2}^r\hat{\kappa},
  \hat{g}_{s1}^r\big]\big) \notag \\
  =& 4 \big[ f_{0,s2} \bar{f}_{0,s1} - f_{0,s1} \bar{f}_{0,s2}  
  - \bm{f}_{s1} \cdot \bar{\bm{f}}_{s2} + \bm{f}_{s2} \cdot \bar{\bm{f}}_{s1} \notag \\
  &+ 2(\bm{m}\cdot\bm{f}_{s1})(\bm{m}\cdot\bar{\bm{f}}_{s2}) 
   - 2(\bm{m}\cdot\bm{f}_{s2})(\bm{m}\cdot\bar{\bm{f}}_{s1}) \big].
\end{align}

Below, we consider the case where the magnetization of the barrier points out of plane
with ${\bm m}=(0,0,1)$. In the clean limit, we have
\begin{align} \label{kgg}
  & {\rm Re}\big[{\rm tr}\big(\tau_3 \sigma_0\big[ \{\hat{\kappa}, \hat{g}_{s2}^r\}, 
    \hat{g}_{s1}^r\big]\big)\big] \notag \\
  =& -16\, {\rm Im}(a^2) (\varepsilon^2 + \beta_{\rm so}^2) J^2 \sin\theta \cos\phi 
  + \cdots ,
\end{align}
and 
\begin{align} 
  & {\rm Re}\big[{\rm tr}\big(\tau_3\sigma_0\big[ \hat{\kappa}\hat{g}_{s2}^r\hat{\kappa},
  \hat{g}_{s1}^r\big]\big)\big] \notag \\
  =& 8\, {\rm Im}\big[ f_{0}\bar{f}_{0} - a^2 (\varepsilon^2 + \beta_{\rm so}^2)
  J^2 \cos\theta \big] \sin\phi .
\end{align}
The terms which are odd in valley index in Eq.~\eqref{kgg} are not shown. 
They do not contribute to the supercurrents since there is a sum over both valleys.
Then the charge current is obtained as
\begin{equation}
  I_c = \big(\sqrt{1-{\cal P}^2} I_{c0} + I_1 \cos\theta \big) \sin\phi 
  + {\cal P} I_1 \sin\theta \cos\phi ,
\end{equation}
as provided in Eq.~(12) in the main text. 

For the spin current in superconductor $s1$, we have
\begin{align}
  & {\rm tr}\big( \tau_3\sigma_z\big[ \{\hat{\kappa}, \hat{g}_{s2}^r\}, 
  \hat{g}_{s1}^r\big]\big)
  = 4 \big[ ig_{0,s2}{\bm m}\times{\bm g}_{-,s1} \notag \\
  &\quad +2({\bm m}\times\bar{\bm f}_{s2}) \times {\bm f}_{s1}
         -2({\bm m}\times{\bm f}_{s2}) \times \bar{\bm f}_{s1} \big]_z \notag \\
  =& 8 \big[ ({\bm m}\times\bar{\bm f}_{s2}) \times {\bm f}_{s1}
            -({\bm m}\times{\bm f}_{s2}) \times \bar{\bm f}_{s1} \big]_z ,
\end{align}
and 
\begin{align}
  & {\rm tr}\big(\tau_3\sigma_z\big[ \hat{\kappa}\hat{g}_{s2}\hat{\kappa}, \
  \hat{g}_{s1}\big]\big) \notag \\
  =& 4i \big[ -2({\bm m}\cdot{\bm f}_{s2}) ({\bm m}\times\bar{\bm f}_{s1})
     -2({\bm m}\cdot\bar{\bm f}_{s2}) ({\bm m}\times{\bm f}_{s1}) \notag \\
  & \quad + 2\bm{g}_{+,s2} \times {\bm g}_{+,s1} +
  \bm{f}_{s2} \times \bar{\bm f}_{s1} - {\bm f}_{s1} \times \bar{\bm f}_{s2} \big]_z 
  \notag \\
  = & {\rm tr}\big(\tau_3\sigma_z\big[ \hat{g}_{s2}, \ \hat{g}_{s1}\big]\big) ,
\end{align}
where the second equality in each of the two equations above are obtained using 
${\bm m}=(0,0,1)$. 
Since
\begin{align} 
  & {\rm Re}\big[{\rm tr}\big(\tau_3 \sigma_z\big[ \{\hat{\kappa}, \hat{g}_{s2}^r\}, 
    \hat{g}_{s1}^r\big]\big)\big] \notag \\
  =& -16\, {\rm Im}(a^2) (\varepsilon^2 + \beta_{\rm so}^2) J^2 \cos\theta \sin\phi
  +\cdots ,
\end{align}
where the terms which are odd in valley index are not shown, 
the spin current is expressed as
\begin{equation}
  I_s = (\hbar/2e)\big[(I_{s0} +  I_1 \cos\phi) \sin\theta
    + {\cal P} I_1 \sin\phi \cos\theta \big].
\end{equation}
as provided in Eq.~(12) in the main text. 

\end{document}